# How to Fit simply Soil Mechanics Behaviour with Incremental Modelling and to Describe Drained Cyclic Behaviours

## P. Evesque

Lab MSSMat, UMR 8579 CNRS, Ecole Centrale Paris
92295 CHATENAY-MALABRY, France, e-mail evesque@mssmat.ecp.fr

**Abstract:**
*It has been proposed recently a new incremental modelling to describe the mechanics of soil. It de. It is based on two parameters called the pseudo Young modulus $E=1/C_o$ and the pseudo Poisson coefficient **n** which both evolve during compression. Evolution of **n** is known since it shall fit the Rowe's law of dilatancy, but $C_o$ has to be evaluated from experiment. In this paper we proposed a way to evaluate the $C_o$ variation from other mechanical modelling. The way cyclic behaviour of drained sample can be modelled is also described.*

______________________________________________________________________________________

We have shown in a series of recent papers [1-4] that classical soil mechanics results corresponding to both oedometric tests and constant volume tests can be described simply with an incremental modelling with a pseudo Poisson coefficient ν and a pseudo Young modulus $E=1/C_o$ as far as the Pseudo Poisson coefficient fits the Rowe's law so that ν depends on the stress ratio and obeys in the notations used in [1-4]:

$$\nu = \sigma_1/[2\sigma_3(1+M)] \qquad (1)$$

with the incremental law:

$$\begin{pmatrix} d\varepsilon_1 \\ d\varepsilon_2 \\ d\varepsilon_3 \end{pmatrix} = -C_o \begin{pmatrix} 1 & -\nu & -\nu \\ -\nu & 1 & -\nu \\ -\nu & -\nu & 1 \end{pmatrix} \begin{pmatrix} d\sigma_1 \\ d\sigma_2 \\ d\sigma_3 \end{pmatrix} \qquad (2)$$

As far as experiments concern oedometric and undrained behaviours and as far as one does not want to predict the evolution of the vertical deformation $\varepsilon_1$, there is no need to determine the evolution of $C_o$ with the stress field. However, for other experimental cases or when knowledge of $\varepsilon_1$ is required one shall get an estimate of the evolution of $C_o=1/E$. A way to estimate this pseudo Young modulus is to use a modelling compatible with classical soil mechanics results.

So the way which is proposed here is to identify a good modelling of soil mechanics and to identify the variation of $C_o$ on a classical triaxial path at $\sigma_2 = \sigma_3 = c^{ste}$. In this case, the identification leads to:

$$C_o = -(\delta\varepsilon_1/\delta\sigma_1)_{\sigma_2=cste} \qquad (3)$$

Eq. (3) is only concerned with plastic deformation, since elastic one is negligible.

We shall now describe a classical model.

**Hujeux modelling:**
We choose the Hujeux modelling [5] , since it is able to describe most of the mechanical features of soils; in particular it can describe the irreversible (*i.e.* plastic) evolution from contractant to dilatant nature of dense soil as deformation proceeds and it respects the classical evolution of the critical state density with pressure [6]:

$$v_c = v_{co} - \lambda \ln(p/p_o) \qquad \text{at large deformation} \qquad (4)$$

where $\lambda$ is a coefficient, $v_{co}$ is the critical volume at $p=p_o$. In order to do so, Hujeux modelling contains two hardening mechanisms, one which is isotropic and the second which is deviatoric. Its load function is given by:

$$F = q/p - M' [1 - b \ln(p/p_{co}) - (b/\lambda)\varepsilon_{v,p}] \varepsilon_{d,p}/(a+\varepsilon_{d,p}) = 0 \qquad (5)$$

where M' is related to the friction angle $\varphi$ (*i.e.* $\sin\varphi = 3M'/(6+M')$; M' is the asymptotic value of the q/p ratio obtained at large deformation during a triaxial compression at constant lateral pressure; b and a are two parameters. Notations are as follows: $\varepsilon_{v,p}$ and $\varepsilon_{d,p}$ denotes the isotropic and deviatoric parts of the plastic deformation; $\varepsilon_{v,e}$ and $\varepsilon_{d,e}$ the isotropic and deviatoric parts of the elastic deformation and the total deformation is the sum of the elastic and plastic deformations. Deformation e will be considered as positive when the volume expands and the sample height increases; in this case a shall be negative. Relations linking the deviatoric and isotropic parts of the deformation and of the stress tensors to the components of the same tensors expressed in principal directions are recalled in table 1:

| $\varepsilon_d = \varepsilon_{d,tot} = \varepsilon_{d,e} + \varepsilon_{d,p}$ | | $\varepsilon_v = \varepsilon_{v,tot} = \varepsilon_{v,e} + \delta\varepsilon_{v,p}$ | |
|---|---|---|---|
| $\varepsilon_1 = [2\varepsilon_d + \varepsilon_v]/3$ | $\varepsilon_2 = \varepsilon_3 = [\varepsilon_v - \varepsilon_d]/3$ | $\varepsilon_d = \varepsilon_1 - \varepsilon_2$ | $\varepsilon_v = \varepsilon_1 + 2\varepsilon_3$ |
| $\sigma_1 = 2q/3 + p$ | $\sigma_2 = \sigma_3 = p - q/3$ | $q = \sigma_1 - \sigma_2$ | $p = [\sigma_1 + 2\sigma_2]/3$ |

***Table 1:*** *Relations between total, elastic and plastic deformation (first line). Relations between the deviatoric and isotropic part of the deformation and the deformation along the principal directions (second line). Relations between the deviatoric and isotropic part of the stress tensor and the stress along the principal directions (third line).*

Typical ranges of parameters a, b, $\lambda$, M, $\varphi$ for different soils are summed up in Table 2.

|  | sands | clays |
|---|---|---|
| a | -0.03=-3% | -.03 to |
| b | 0.12 - 0.2 | 1 |
| M'=q/p | 1,2 | 1.2 |
| $\varphi$ | 30°-40° | 30° |
| $\lambda$ | 0.06 | 0.1 |

***Table 2:*** *Typical values of the parameters entering Hujeux modelling.*

As mentioned previously, M' is related to the friction angle of the critical state since $\sin\varphi = 3M'/(6+M')$. Parameter a controls the deviatoric deformation at which the change from contractant to dilatant behaviours occurs. $\lambda$ controls the evolution of the critical volume with mean pressure p, see Eq. (4); hence, $p_{co}$ characterises the initial density of the material; it is called the overconsolidation pressure when soil is clay. According to Eq. (4) the larger $p_{co}$, the larger the initial density and the smaller the initial specific volume $v_{co}$. Endly, b controls the amplitude of the peak of deviatoric stress.

Hujeux modelling obeys classical plasticity theory. So, it requires to introduce a flow rule which governs the plastic deformation of the material and to introduce the elastic Young modulus $E_e$ and the elastic Poisson coefficient $\nu_e$, which control the elastic part of the deformation. It is worth recalling that $E_e$ and $\nu_e$ have not to be confused with the pseudo Young modulus $E=1/C_o$ we are looking for and with the pseudo Poisson coefficient $\nu$ given by Eq. (1). These coefficients $E_e$ and $\nu_e$ are assumed to be constant. This leads to the three new equations:

$$q = \varepsilon_{d,e}[E_e/(1+\nu_e)] \quad <=> \quad \varepsilon_{d,e} = [(1+\nu_e)/E_e]\, q \quad (6)$$

$$p = \varepsilon_{v,e}[E_e/\{3(1-2\nu_e)\}] \quad <=> \quad \varepsilon_{v,e} = [3(1-2\nu_e)/E_e]\, p \quad (7)$$

$$\delta\varepsilon_{v,p} = \delta\varepsilon_{d,p}\,(M-q/p) \quad (8)$$

Eq. (8) is the plastic flow rule introduced by Roscoe [6]. In the next, we will assume the elastic part of deformation to be negligible or, what is the same, $E_e \to \infty$ and $\nu_e$ finite.

## *Finding $C_o$:*

Let us consider a sample submitted to a stress field $(\sigma_1,\sigma_2=\sigma_3)$ or (q,p), which has been deformed $(\varepsilon_1,\varepsilon_2=\varepsilon_3)$ or $(\varepsilon_v,\varepsilon_d)$. The problem is now to determine the parameters entering Eq. (2) and there evolution with deformation and stress, *i.e.* $1/C_o$ and $\nu$. $\nu$ is given by Eq. (1) (or by Eq. 8). In order to find $C_o$, one can perform an incremental compression at $\sigma_2=\sigma_3=c^{ste}$ and identify $C_o$ from the comparison of Eq. (2) and of the experimental result. An other way to proceed is to assume that previous modelling such as Hujeux modelling is correct and can be used to compute the experimental response.

This is just what we will do: As the elastic part of the deformation is quite small we can neglect it and identify the total deformation and the plastic one. As the deformation is plastic, the load function impose dF=0 during the deformation. This leads to a relation between $\delta q$, $\delta p$, $\delta\varepsilon_{d,p}=\delta\varepsilon_d$ and $\delta\varepsilon_{v,p}=\delta\varepsilon_v$. But, since $\delta\sigma_2=\delta\sigma_3=0$, one gets $\delta q=\delta p/3$ (see Table 1), and $\delta\varepsilon_v=\delta\varepsilon_d\,(M-q/p)$ owing to Eq. (8). So writing dF=0 allows to find $\delta\varepsilon_1$, $\delta\varepsilon_2=\delta\varepsilon_3$, for a given $\delta\sigma_1$. And to find $C_o=\delta\varepsilon_1/\delta\sigma_1$:

$$-C_o(1+\nu) = \left[ 1 - (1/3)\left\{\{M(1-b) - Mb\,\ln(p/p_{co}) - (Mb/\lambda)\,\varepsilon_v\}[\varepsilon_{d,p}/(a+\varepsilon_d)]\right\}\right] \Big/ \Big\{ (Mp-q)(Mb/\lambda)[\varepsilon_d/(a+\varepsilon_d)] + [a/(a+\varepsilon_d)^2] \ast$$

$$\{Mp - Mbp \ln(p/p_{co}) - (Mb/\lambda) p \, \varepsilon_{v,p}\}\} \quad (9a)$$

with  $\nu = \sigma_1/[2\sigma_3(1+M)] = (2q+3p)/[(6p-2q)(1+M)]$ (9b)

It is worth noting that $C_o$ is approximately inversely proportional to the applied stress field ($\sigma_2$). It is worth noting that the incremental response depends on the stress field but also on the deformation; it is then depending on the path. This complicates the prediction, but it is coherent with what one knows of soil behaviour.

### *Typical compression behaviours obtained with this modelling and at $\mathbf{s}_2 = c^{ste}$*

Having determined $C_o$ by this method, one can integrate the evolution along a classical compression path at $\sigma_2$ constant using Eq. (1) and (2). Typical examples are reported in Fig. 1, where the evolution of q/p , of $v/v_o$ and of the pseudo Young modulus are plotted as a function of $(h_o-h)/h_o = -\varepsilon_1$ for different values of initial "overconsolidated pressure" $p_{co}$. As one expects, the results depend effectively on the initial specific volume (or the initial pressure of overconsolidation $p_{co}$) compared to the critical one; the denser the initial sample, i.e. the larger $p_{co}$, and the smaller $\sigma_2$, the larger the maximum of q/p , the larger the dilatancy effect and the larger the volume increase. For small $p_{co}$, i.e. $p_{co}=\sigma_2$, the volume is always contracting all along the compression, and the q/p ratio deviatoric stress increases till it reaches M'. Values of $p_{co}$ smaller than $\sigma_2$ is not physical, but can be used. Simulations confirm that $1/C_o$ (or E) is approximately proportional to the stress field ($\sigma_2$), the evolution of q/p occurs always in the same range of $\varepsilon_1$, although the precise evolution depends on the precise value of $p_{co}$ for a given p .

One remarks also that when the overconsolidation pressure $p_{co}$ is quite large, the pseudo Young modulus E increases at small deformation. This is because the sample contracts during the first part of deformation. This volume decrease strengthens the material ; the E increase is relatively smaller when $\sigma_2$ is larger (for the same $p_{co}$) because the value of E is much larger since it is proportional to $\sigma_2$ ; so the relative change $\delta E/E$ is smaller. This effect has not been mentioned by previous studies [5] of Hujeux modelling ; it is probably because this effect is small and washed out by the elasticity term which is incorporate in the model.

One can understand how the modelling leads to the mechanical behaviour reported in Fig. 1: at large deviatoric deformation, *i.e.* $|\varepsilon_d|$»$|a|$, the term $\varepsilon_d/(a+\varepsilon_d)$ remains constant, so that the system obeys the load function of the Granta gravel [6] and depends of the specific volume v; so, Hujeux modelling is equivalent to Granta gravel when $|\varepsilon_d|$»$|a|$. However, when it becomes small, *i.e.* $|\varepsilon_d|$«$|a|$, the term $d\{\varepsilon_d/(a+\varepsilon_d)\}$ dominates dF, modifying the behaviour predicted by Granta gravel imposing a deformation process rather independent of v.

According to this explanation, it is now easy to modify Hujeux modelling if one desires a non linear variation of q/p vs. $\varepsilon_1$ in the small $\varepsilon_1$ regime, *i.e.* $|\varepsilon_d|$ «$|a|$ : One has just to change the term $\varepsilon_d/(a+\varepsilon_d)$ in F ,i.e. Eq. (5), by a term $\varepsilon_d^n/(a^n+\varepsilon_d^n)$. In this case, however, $C_o$ will not be inversely proportional to the stress field (q,p), and the range of interest for the deformation $\varepsilon_l$ will depend on the stress level; it will vary with p.

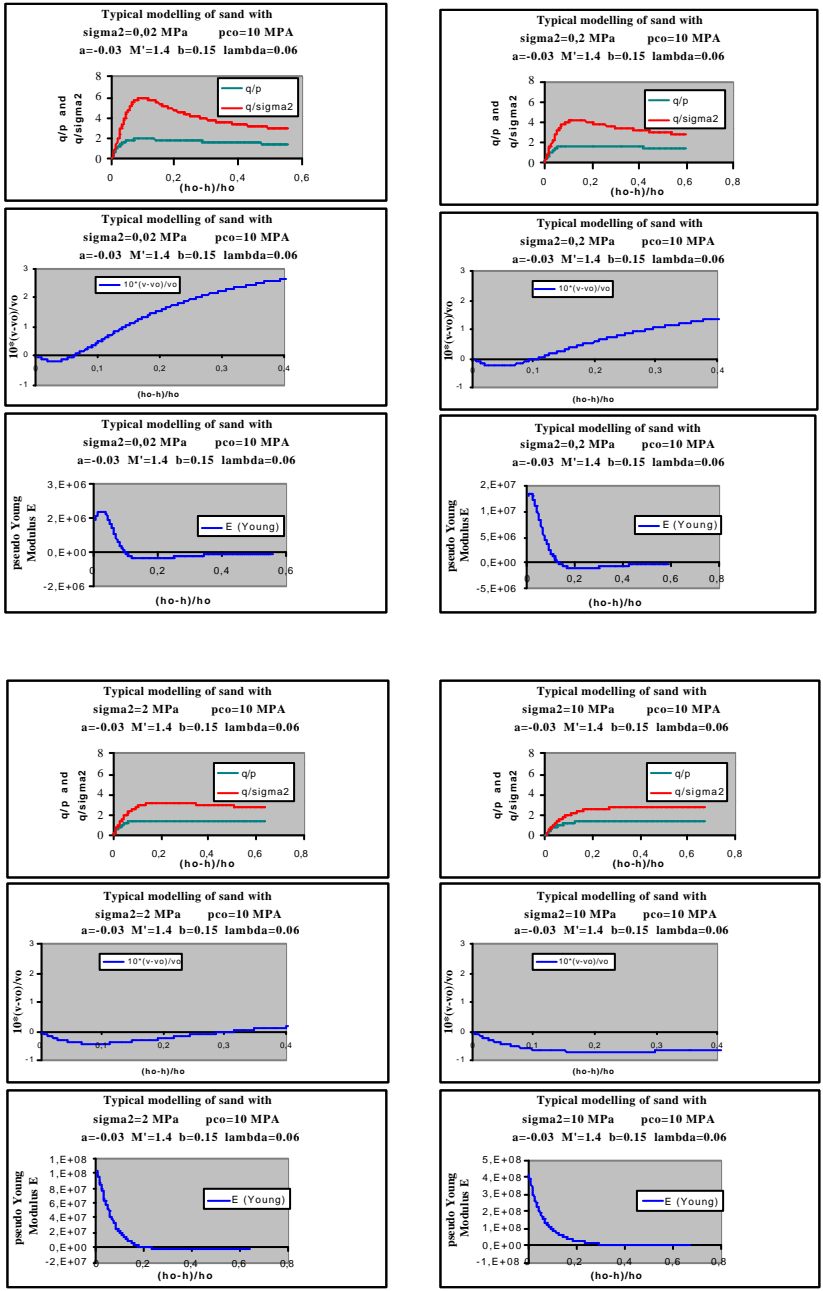

***Figure 1:*** *Typical stress-strain behaviours of sand predicted by the Hujeux model, for different set of parameters (see text). Parameters*: a=-0.03 ; M'=1.4 ; b=0.15 ; λ=0.06; $p_\infty$=100Mpa

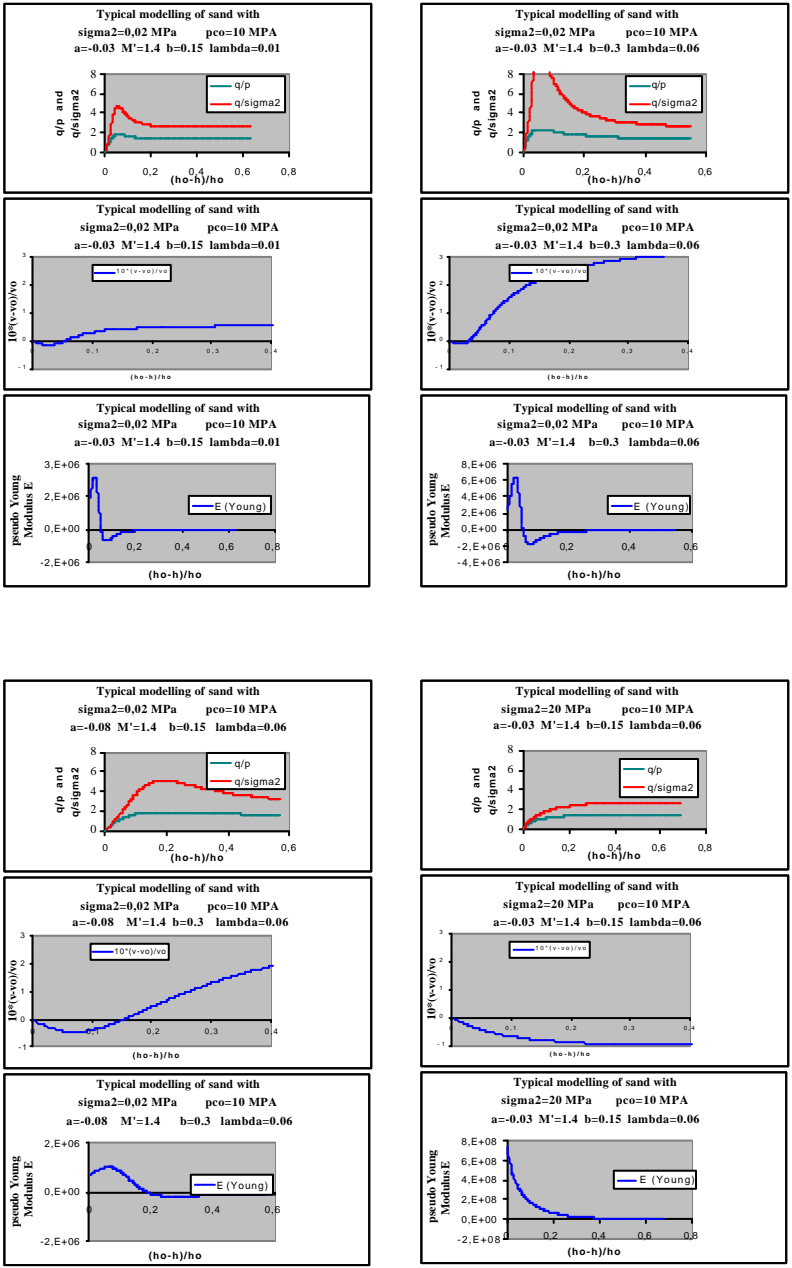

**Figure 2:** *Typical behaviours predicted by Hujeux modelling for two other sets of parameters: Left-top: effect of* b=0.3 ; *Right-top: effect of* λ=0.01 ; *Left-bottom: effect of* a=-0.08 ; *Right-bottom:* σ₂ < p_ω .

### *Range of validity of parameters* a, b, **l**, $p_{co}$:

Hujeux model is sensitive to the values of the parameters. Typical examples of variation of behaviour are reported in Fig. 2. Unphysical predictions can also be obtained, since one can use s2 values larger than the overconsolidation values, as shown in Fig.2. Also the range of deformation can be extended to the real one of interest ($\varepsilon$=0-1).

### **Describing cyclic behaviours:**

### *Experimental observation:*

It is known that cyclic behaviours of soil are mainly controlled by
(i)     the average value <q> at which the cycles are performed,
(ii)    the amplitude $\Delta q$ of the cycles.

In particular, the contractant vs. dilatant nature of the average effect which is obtained after each cycle depends on the value of <q>/p compared to M': if <q>/p >
 *i.e.* $\Delta v>0$ , otherwise, it contracts, *i.e.* $\Delta v<0$.

Secondly, it is known also that this mean effect depends on, and decreases with, the number N of cycles already performed. It evolves with the number N of cycle in a logarithmic way and scales as:

$\Delta_N v = \Delta_1 v / \ln(N)$ (10)

Thirdly, the larger $\Delta q/p$ the larger the volume variation.

Fourthly, one knows also that the value of <q> - $\Delta q$ is quite important; in particular it is known that a change of sign of q during the cycle, involving an interchange of the axis of compression fom z to (x,y) increases strongly the cycle effect.

When the experiment is run at constant volume, *i.e.* undrained condition, cycles performed at <q>/p > M' imposes a decrease of water pressure $u_w$ in the pore and an increase of p, whereas cycles performed at <q>/p <M' generates at each cycle an increase of the water pressure $u_w$ in the pore and a decrease of p. These two phenomena stop when the value of <q>p reaches the M' value, i.e. the characteristic-state line [7]. In the case when <q>= 0, the total mean pressure $p_\infty$ exerted on the grain skeleton which is reached after a great number of cycles is then 0, so that the medium is no more able to sustain shear so that it is liquefied. This is the liquefaction process.

It is known that the true Hujeux modelling [5], which incorporates elastic and plastic deformation, is able to describe these phenomena [5]. It is possible also our modelling for drained test with. It requires simply to change the hardening parameter $\varepsilon_{d,p}$ in F, *i.e.* Eq. (5), at each time; but we will describe this point a little later.

### *The evolution of the pseudo Poisson coefficient with stress field ensures the compatibility with experimental data on drained samples:*

Let us first show that the proposed modelling is compatible with the fact that each cycle leads to compaction or dilatation depending on the value of the <q>/p ratio; for this, we limit the analysis to a cyclic uniaxial compression test performed at $\sigma_2=\sigma_3=c^{ste}$ on a perfectly drained sample: During half a period the sample is deforming plastically; during the other half it behaves "elastically", however as $E_e\to\infty$, elastic deformation remains small during this second halve of period so that the total deformation during one cycle is about the one obtained during the first half period. The contractant/dilatant nature of each cycle is controlled by the mean value of the pseudo Poisson coefficient $\nu$ during the first halve of cycle, i.e. $\Delta v=-\int 3C_o(1-2\nu)dp$ from Eq. (2), with $dp=d\sigma_1/3$. Hence, in the case of small cycles, one concludes that (i)

$\Delta v >0$ if $\nu>1/2$   *i.e.* for $<q>/p>M'$   (11a)

$\Delta v <0$ if $\nu <1/2$   *i.e.* for $<q>/p<M'$   (11b)

which is just what is observed.

### *How to model the evolution of the drained cyclic behaviour with the proposed modelling:*

So, most of the features of cyclic behaviour is described due to the evolution of $\nu$ with the stress field described by Eq. (1); it shall also take into account the irreversible plastic deformation which is generated during half part of the cycle. However this is not sufficient if one wants to model the variation of the cycle effect with the number of cycles .

In order to model this effect of cycle number, it is required to introduce a hardening effect in F, *i.e.* Eq. (5), which will diminish the amplitude of variation due to each cycle as the number of cycle proceeds. This hardening can be controlled by the value $\varepsilon_d$ which has to be taken at the beginning of each cycle, or by making evolving the value of the parameter a,… with N in Eq. (5) if one wants to take into account the number N of cycles, or to divide the effective value taken by $C_o$ in Eq. (2) by Ln(N). Furthermore, when cycles are such that q passes through 0 (alternate cycle) then $\varepsilon_d$ can be set to 0 at the beginning of each next cycles, in order to cancel the memory effect; if cycles do not passes through q=0 but remains always positive then $\varepsilon_d$ can be changed in $\varepsilon_d -\Delta\varepsilon_d$ in Eq. (5); in this case, the value of $\Delta\varepsilon_d$ has to depend on the relative cycle amplitude and on the number N of cycles already performed; in order to fit the Ln(N) effect, this means: $\Delta\varepsilon_d =f(\Delta q/q)/Ln(N)$ .

*Acknowledgements:* I thank A. Modaressi for quite stimulating discusssions.

The electronic arXiv.org version of this paper has been settled during a stay at the Kavli Institute of Theoretical Physics of the University of California at Santa Barbara (KITP-UCSB), in june 2005, supported in part by the National Science Fundation under Grant n° PHY99-07949.

*Poudres & Grains* can be found at :
http://www.mssmat.ecp.fr/rubrique.php3?id_rubrique=402